\documentclass[letter, reqno]{amsart}
\usepackage[margin=1.1in]{geometry}
\usepackage{amsmath}
\usepackage{graphicx}
\usepackage{enumerate}
\usepackage{natbib}
\usepackage{url}
\usepackage{amssymb}
\usepackage[foot]{amsaddr}
\usepackage{setspace}
\usepackage{mathtools}
\usepackage{bbm}
\usepackage{booktabs}
\usepackage{amsthm}
\usepackage{algorithm}
\usepackage{algpseudocode}
\usepackage{pdfpages}

\theoremstyle{definition}
\newtheorem{assumption}{Assumption}

\newtheorem{theorem}{Theorem}
\newtheorem{corollary}{Corollary}

\newcommand{\norm}[1]{\left\Vert#1\right\Vert}
\newcommand{\abs}[1]{\left\vert#1\right\vert}

\def\eps{\varepsilon}

\def\argmin{\mathop{\rm argmin}}

\def\Ind{\mathbb{I}}

\def\E{\mathbb{E}}
\def\Prob{\mathbb{P}}
\def\V{\mathbb{V}}

\def\R{\mathbb{R}}
\def\Z{\mathcal{Z}}
\def\X{\mathcal{X}}

\def\Ncal{\mathcal{N}}
\def\Fcal{\mathcal{F}}
\def\Scal{\mathcal{S}}
\def\Ecal{\mathcal{E}}
\def\Bcal{\mathcal{B}}
\def\Acal{\mathcal{A}}

\def\Zcal{\mathcal{Z}}

\def\sumi{\sum_{i=1}^N}
\def\sumj{\sum_{j=1}^N}

\def\partz{\frac{\partial}{\partial z}}

\def\Lj{L_{z_0}(j)}

\def\xv{\mathbf x}

\def\zv{\mathbf z}

\def\Av{\mathbf A}

\def\Uv{\mathbf U}

\def\Xv{\mathbf X}
\def\Yv{\mathbf Y}
\def\Zv{\mathbf Z}

\title[Local Causal Effects with Continuous Exposures]{Local Causal Effects with Continuous Exposures: A Matching Estimator for the Average Causal Derivative Effect}

\author{Suhwan Bong$^{1}$}
\address{$^{1}$Department of Statistics, Seoul National University, Seoul, Republic of Korea}
\author{Kwonsang Lee$^{1\ast}$}
\address{$^{\ast}$To whom correspondence should be addressed}


\begin{document}
\maketitle

\begin{abstract}
{
The estimation of causal effects is a fundamental goal in the field of causal inference. However, it is challenging for various reasons. One reason is that the exposure (or treatment) is naturally continuous in many real-world scenarios. When dealing with continuous exposure, dichotomizing the exposure variable based on a pre-defined threshold may result in a biased understanding of causal relationships. In this paper, we propose a novel causal inference framework that can measure the causal effect of continuous exposure. We define the expectation of a derivative of potential outcomes at a specific exposure level as the average causal derivative effect. Additionally, we propose a matching method for this estimator and propose a permutation approach to test the hypothesis of no local causal effect. We also investigate the asymptotic properties of the proposed estimator and examine its performance through simulation studies. Finally, we apply this causal framework in a real data example of Chronic Obstructive Pulmonary Disease (COPD) patients.
}
\end{abstract}

\setstretch{2}

\section{Introduction}\label{sec1}
The estimation of causal effects is a fundamental goal in the field of causal inference. However, it is often challenging for various reasons. First, randomized experiments may not always be feasible due to ethical concerns or high costs, leaving researchers to rely on observational studies that are susceptible to the impact of potential confounders that can bias the estimation of causal effects. Second, the exposure (or treatment) is naturally continuous in many real-world scenarios, posing a significant challenge for researchers. The complex nature of continuous exposures can make it difficult to derive reliable estimates of causal effects. Third, even a considerable number of existing estimation methods rely on parametric or nonparametric models, which can be sensitive to model misspecification.

When dealing with continuous exposure, dichotomizing the exposure variable based on a pre-defined threshold may result in a biased understanding of causal relationships. Therefore, identifying the causal effect of continuous exposure requires an appropriate causal framework that can adequately account for the continuous nature of the exposure variable. The propensity score method, initially introduced by \cite{rosenbaum1983central} for binary exposure variables, has been extended to continuous exposure variables \citep{hirano2004propensity, imai2004causal} to adjust for confounding. One approach is to include the estimated generalized propensity score (GPS) as a covariate in the outcome model \citep{hirano2004propensity, kluve2012evaluating}. However, the validity of this method relies on the correct specification of both the exposure and outcome models. Another method, the inverse probability of treatment weighting (IPTW) approach, was proposed by \cite{robins2000marginal} to adjust for confounding in continuous exposures. However, IPTW can suffer from instability due to high weights assigned to certain individuals. Moreover, the true GPS may be unknown in practice, making it necessary to rely on specific models for estimation. To address these issues, several methods have been proposed that directly optimize weights without relying on explicit models to estimate the GPS, such as the covariate balancing generalized propensity score (CBGPS) approach \citep{fong2018covariate}, the covariate association eliminating weights \citep{yiu2018covariate}, and the entropy balancing weights \citep{vegetabile2021nonparametric}. Additionally, some methods focus on flexible modeling of the GPS and outcome models, such as boosting and nonparametric methods \citep{zhu2015boosting, kennedy2017non}. However, these methods all rely on the estimation of the GPS and sensitive to extreme values of the estimated GPS \citep{wu2022matching}. Assessing covariate balance and conducting sensitivity analyses for unmeasured confounders can be challenging in these approaches, which can limit their practical applicability.

Matching methods have been proposed to address the limitations of previous methods in estimating causal effects, and these methods have been well-established in the literature \citep{rubin1973matching, rosenbaum1989optimal, gu1993comparison, heckman1998matching, abadie2002simple}. Matching methods are robust to model misspecification and extreme values of the estimated weights. However, most existing matching methods are limited to binary exposure variables. Among them, there are only a few studies on matching methods that can be used with continuous exposure variables \citep{zhang2023statistical, wu2022matching}. \cite{zhang2023statistical} divided the original set via non-bipartite matching method and fitted a regression model to each group. Also, \cite{wu2022matching} matched a dataset on GPS and estimate a exposure-response function. However, none of them addressed the local impact of an exposure on the outcome. Moreover, to the best of our knowledge, hypothesis test or sensitivity analysis for unmeasured confounders has not been performed for matching methods with continuous exposures. The presence of unmeasured confounders may lead to biased estimates of causal effects and threaten the validity of causal inference.

While previous works on continuous exposures have mostly targeted the population exposure-response function (ERF) as a causal estimand, other estimands may also be of interest depending on the research question. To measure the causal effect of a binary exposure on an outcome, we typically calculate the expected outcome differences between two interventions, such as the average treatment effect ($\E[Y(1)] - \E[Y(0)]$), rather than an expected outcome itself ($\E[Y(1)]$). Similarly, in the case of a continuous exposure, we may be interested in the expected outcome difference resulting from a small change in exposure, such as $\E[Y(z+h)] - \E[Y(z)]$, for small $h$. This estimand can explain the causal effect of exposure on an outcome when there is a small perturbation in the target exposure level. If we further assume some regularity conditions, this difference can be estimated as $h\cdot\E[Y'(z)]$. We denote $\E[Y'(z)]$ as the causal derivative effect at $z$. While this serves as an instance of a weighted average derivative estimator \citep{powell1989semiparametric}, prior studies focusing on this estimator have primarily employed semiparametric methods without addressing confounder variables \citep{powell1989semiparametric, hardle1989investigating, cattaneo2010robust}. In our paper, we utilize a derivative estimator within a causal framework. This causal estimand is applicable in various scenarios where assessing the local effect of an exposure in proximity to a specific exposure level holds more significance than measuring the average response at that exposure level itself.

The paper is structured as follows. In Section~\ref{sec2}, we introduce the notation and causal inference framework essential for identifying the causal estimand of interest, the average causal derivative effect. Section~\ref{sec3} outlines the matching algorithm and the data-driven hyperparameter selection method employed for estimating the average causal derivative effect. Section~\ref{sec4} introduces a permutation test approach to assess the hypothesis of no local exposure effect and discusses sensitivity analysis for unmeasured confounders. Additionally, in Section~\ref{sec5}, we present the asymptotic properties of our matching estimator, including its consistency and asymptotic normality under certain assumptions. We examine the proposed properties through simulation studies in Sections~\ref{sec4} and \ref{sec5}. Furthermore, we apply our methodology to COPD patient data in Section~\ref{sec6}. Finally, we conclude the paper with a discussion in Section~\ref{sec7}.

\section{Notation and Causal Framework}\label{sec2}

In this section, we introduce the notation and causal framework used in this paper to estimate the derivative effect of a continuous exposure on an outcome. We consider a continuous exposure $Z$ with bounded support $\Z \subset \R$. Assume that there are $N$ individuals. For each individual $i = 1,2,\dots,N$, we observe the exposure level $Z_i \in \Z$ and the covariate $\Xv_i \in \X \subset \R^d$, where $\Xv_i = (X_{i1}, X_{i2}, \dots, X_{id})^\top$ is a covariate vector of dimension $d$. Let $Y_i(z)$ denotes the potential outcome of an individual $i$ at the exposure level $z$. We assume that the potential outcome $Y_i(z)$ changes continuously as the exposure level $z$ changes for each $i$. We only observe one of the potential outcomes for each $i$, denoted as $Y_i$. In observational studies, exposure levels are never randomly assigned to individuals, making it difficult to establish causal relationships between exposure and outcome. Confounding bias arises due to the fact that the exposure level $Z_i$ and the outcome $Y_i$ are both influenced by the potential confounder $\Xv_i$. We adopt the potential outcome framework with continuous exposures introduced by \cite{hirano2004propensity} and establish the following assumptions to identify the causal estimand of interest.

\begin{assumption}[Consistency]\label{ass1}
    For each $i$, $Z_i = z$ implies $Y_i = Y_i(z)$.
\end{assumption}

Assumption~\ref{ass1} implies that potential outcomes are determined by an individual's exposure level while not influenced by the exposure levels of others. Thus, we observe one of the potential outcomes, $Y_i = Y_i(Z_i)$ for individual $i$. Also, \cite{hirano2004propensity} introduced a weak confoundedness assumption, $Y_i(z) \perp\!\!\!\perp Z_i \vert \Xv_i$ for all $z\in\Z$. Many works dealing with continuous exposure are based on this assumption \citep{galvao2015uniformly, kennedy2019nonparametric, kluve2012evaluating, kreif2015evaluation, vegetabile2021nonparametric}. However, \cite{wu2022matching} introduced local weak confoundedness assumption to estimate the exposure-response function. Since we are interested in measuring the causal effect of an exposure nearby $Z=z_0$ to the outcome, the local weak confoundedness assumption allows us to identify the causal effect. We define the caliper $\eta > 0$, which is used as the radius of the neighborhood for a certain exposure level $z_0$. We consider $\eta$ as a constant that decreases to 0 as a sample size $N$ goes to infinity.

\begin{assumption}[Local Weak Confoundedness]\label{ass2}
    For all $z\in\Z$, $f(Y_i(z)\vert Z_i = \Tilde{z}, \Xv_i) = f(Y_i(z) \vert \Xv_i)$ for any $\Tilde{z}\in [z-\eta, z+\eta]$ where $f$ denotes a generic probability density function.
\end{assumption}

Also, we need to guarantee that every individual has some chance of assigning exposure level $z\in\Z$; thus, the exposure level assignment is not deterministic. We now introduce Assumption~\ref{ass3}, the Strict Overlap assumption.

\begin{assumption}[Strict Overlap]\label{ass3}
    For all $z \in \Z$ and $\xv\in\X$, the conditional probability density function is positive, that is, $f_{Z_i\vert \Xv_i}(z\vert \xv) \geq \zeta$ for some constant $\zeta>0$.
\end{assumption}

Under Assumption~\ref{ass1}-\ref{ass3}, it is possible to identify the exposure-response function $\E[Y_i(z)]$ using observable variables as $\E[\E\{Y_i\vert Z_i = z, \Xv_i\}]$. Now, we will introduce the average causal derivative effect (ACDE) to measure the local exposure effect. We first introduce necessary notations and an additional assumption.

For $z\in\Z$ and $\xv\in\X$, let $\mu(z,\xv) = \E[Y_i\vert Z_i=z, \Xv_i=\xv]$, $\mu_{\xv}(z)=\E[Y_i(z)\vert\Xv_i=\xv]$, $\sigma^2(z,\xv)=\V[Y_i\vert Z_i=z,\Xv_i=\xv]$, $\sigma^2_{\xv}(z)=\V[Y_i(z)\vert\Xv_i=\xv]$, and $\eps_i=Y_i-\mu_{\Xv_i}(Z_i)$. Assumption~\ref{ass2} implies that conditional on covariate $\Xv_i$, an individual's exposure assignment is independent of the potential outcomes. In other words, $\Ind(Z_i=z)\perp\!\!\!\perp Y_i(z)\vert\Xv_i$ holds where $\Ind(\cdot)$ denotes an indicator function. Under Assumption~\ref{ass2}, $\mu_{\xv}(z) = \mu(z,\xv)$ and $\sigma^2_{\xv}(z)=\sigma^2(z,\xv)$ hold. To define the causal estimand of our interest, we also need Assumption~\ref{ass4}.

\begin{assumption}[Smoothness]\label{ass4}
    For each $i$, the potential outcome function $Y_i(z)$ is differentiable with respect to $z$ and its derivative $Y_i'(z) \coloneqq \partz Y_i(z)$ is continuous and bounded for all $z\in\Z$. Also, $\mu_{\Xv_i}(z)$ is differentiable and its derivative $\mu_{\Xv_i}'(z) \coloneqq \partz \mu_{\Xv_i}(z)$ is continuous and bounded for all $z\in\Z$. 
\end{assumption}

We define the ACDE, which can be used to measure the local effect of exposure nearby $z_0$ on outcome, as $\E[Y_i(z_0)]$. Assumption~\ref{ass4} guarantees the differentiability and smoothness of desired functions. Our target causal estimand is the population ACDE at the interested exposure level $z_0 \in \Z$, $\delta(z_0)$.

\begin{align}
    \delta(z_0) \coloneqq \E[Y_i'(z_0)].
\end{align}

This value is different from the incremental causal effect defined by \cite{rothenhausler2019incremental}. They defined the incremental causal effect as an expected change in outcome when we change the current treatment assignment by a small amount, $\E[Y_i'(Z_i)]$, which can be considered as a limit of stochastic interventions. It can explain the expected outcome change if we change the current treatment of all individuals by a small amount \citep{rothenhausler2019incremental}. However, ACDE can explain the expected outcome change if we set the whole population's exposure to $z_0$ and change the exposure level by a small amount. The ACDE is a deterministic way that can account for the local effect of an exposure nearby $z_0$, which is similar to the average treatment effect (ATE) in a binary exposure situation. If $\delta(z_0)$ is significantly greater than 0, we can conclude that the exposure has a critical effect on outcome locally at $Z = z_0$. The following theorem shows that the ACDE can be identified under Assumptions~\ref{ass1}-\ref{ass4}.

\begin{theorem}[Identification of the ACDE]\label{thm1}
    Under Assumptions~\ref{ass1}-\ref{ass4},
    \begin{align*}
        \E[Y_i'(z)] = \partz \E[Y_i(z)]
    \end{align*}
    holds for all $z\in\Z$.
\end{theorem}

Since the exposure-response function is identifiable, we can also identify the ACDE. $\delta(z)$ can be identified as a derivative of the exposure-response function.

\section{Matching Framework}\label{sec3}
Many studies estimated the expected outcome via the regression models \citep{hirano2004propensity, kluve2012evaluating, rothenhausler2019incremental, vegetabile2021nonparametric}. Similarly, if we assume specific parametric or nonparametric regression models on the exposure and outcome, we obtain the ACDE by differentiating the exposure-response function. However, this method is limited in that the underlying model is never known and can be susceptible to model misspecification. Furthermore, even though we estimate the exposure-response function correctly, the hypothesis test for a certain derivative effect or sensitivity analysis is limited. Therefore, we suggest a novel nonparametric matching method for estimating the ACDE, which is flexible and robust compared to traditional regression-based approaches. In addition, this matching method also can naturally be applied to hypothesis testing and sensitivity analysis for unmeasured confounders.

\subsection{Matching Algorithm}\label{sec3.1}\hfill\break
To measure a derivative effect of exposure for an individual $i$ at an exposure level of $Z_i = z_0$, we aim to find pairs of individuals with similar covariates to individual $i$, denoted as $i_1 \equiv i_1(z_0)$ and $i_2 \equiv i_2(z_0)$.  Individual $i_1$ has an exposure level $Z_{i_1}$ slightly less than $z_0$, while individual $i_2$ has an exposure level $Z_{i_2}$ slightly greater than $z_0$. Then, we could estimate the individual derivative effect of $i$ at $z_0$ by comparing the outcomes of these two matched individuals, $Y_{i_1}$ and $Y_{i_2}$. We may estimate the individual derivative effect $\delta_i(z_0)$ as $(Y_{i_2} - Y_{i_1}) / (Z_{i_2} - Z_{i_1})$. Also, we introduce two hyperparameters, denoted as $\eta$ and $\kappa$, to guarantee a reliable estimation of a derivative effect. $\eta$ is used to ensure that the exposure levels of the matched individuals are similar to $z_0$ and $\kappa\in (0,1)$ determines the extent of the difference in exposure levels of the matched individuals. If the difference between $Z_{i_2}$ and $Z_{i_1}$ is extremely small, the estimator for $\delta_i(z_0)$ may become unstable since the denominator is close to zero. The role of these hyperparameters will be further discussed in detail in Section~\ref{sec3.2}.

We now introduce a matching algorithm for estimating the ACDE at $z_0$. For each individual $i = 1,2,\dots, N$, we match the nearest individuals with replacements. The distance between covariates can be measured using any metric $d(\Xv_i, \Xv_{i'})$ for individual $i$ and $i'$. For example, we may choose the standard Euclidean norm, i.e., $d(\Xv_i, \Xv_{i'}) = \norm{\Xv_i - \Xv_{i'}}_2$. Let $i_1$ and $i_2$ be the indices in $\{1,2,\dots,N\}$ satisfying
\begin{align*}
    i_1 = \argmin_{l:z_0-\eta\leq Z_l\leq z_0-\kappa\eta} d(\Xv_l, \Xv_i) \quad\text{and}\quad
    i_2 = \argmin_{l:z_0+\kappa\eta\leq Z_l\leq z_0+\eta} d(\Xv_l, \Xv_i)
\end{align*}
for some pre-defined hyperparameters $\eta>0$ and $\kappa \in (0,1)$. That is, we find an individual $i_1$ whose covariate $\Xv_{i_1}$ is the nearest to $\Xv_i$ with $Z_{i_1} \in [z_0-\eta, z_0-\kappa\eta]$ whereas an individual $i_2$ represents an individual whose covariate $\Xv_{i_2}$ is the nearest to $\Xv_i$ with $Z_{i_2} \in [z_0+\kappa\eta, z_0+\eta]$. $Y_{i_1} (Y_{i_2})$ can be seen as an expected outcome when we change the exposure level of an individual $i$ to slightly less (greater) than $z_0$. If $Y_{i_2}$ is significantly larger than $Y_{i_1}$, this suggests that the exposure has a positive effect on the outcome at $Z_i = z_0$ for individual $i$. The natural estimator for $\delta_i(z_0)$ is given as $\hat\delta_i(z_0) = \frac{Y_{i_2}-Y_{i_1}}{Z_{i_2}-Z_{i_1}}$. Then, the matching estimator of the ACDE at $z_0$, $\hat\delta(z_0)$ can be defined as 
\begin{align}\label{eqn:ACDE}
    \hat\delta(z_0) &= \frac{1}{N}\sumi\hat\delta_i(z_0)\nonumber\\
    &= \frac{1}{N}\sumi\frac{Y_{i_2}-Y_{i_1}}{Z_{i_2}-Z_{i_1}},
\end{align}
which is an average of an estimated individual derivative effect at $z_0$, $\hat\delta_i(z_0)$.

\subsection{Data-driven Hyperparameter Selection and Covariate Balance}\label{sec3.2}\hfill\break
The matching technique helps to control for the effects of possible confounders and improves the validity of the causal inference. However, the performance of the estimator depends on the quality of the matching, i.e., if the covariates of the matched individuals are truly balanced. If the covariates are not balanced, the matching estimator may be biased. We will choose the hyperparameters $(\eta, \kappa)$ based on their ability to balance the covariates, which is crucial in ensuring an accurate estimation result. Selecting a large value for $\eta$ and a small value for $\kappa$ is required to ensure good matching quality. This choice implies a broader range of potential matches with similar covariates. However, if $\eta$ is too large, it will result in possibly increased approximation error and lead to a biased estimator. On the other hand, $\kappa$ should be large enough to ensure that there are significant differences in exposure levels between individuals $i_1$ and $i_2$. If $Z_{i_2}-Z_{i_1}$ is too close to zero, the weights of some $Y_i$ can become too large and dominate the matching estimator. One feasible option for selecting $\kappa$ is to use a rule of thumb, such as 0.1 or 0.01.

In this section, we propose a data-driven approach for determining the optimal hyperparameters $(\eta, \kappa)$. For each individual $i$, we match them with two other individuals, $i_1$ and $i_2$. To ensure the quality of the matching, it is essential that the covariate vectors, $\Xv_i$, $\Xv_{i_1}$, and $\Xv_{i_2}$, are all similar. This means that the original set of covariates $\Xv^\ast=\{\Xv_1, \dots, \Xv_N\}$, as well as the two matched sets $\Xv_1^\ast = \{\Xv_{1_1}, \dots, \Xv_{N_1}\}$ and $\Xv_2^\ast = \{\Xv_{1_2}, \dots, \Xv_{N_2}\}$, should be balance. In the triplet matching method proposed by \cite{nattino2021triplet}, the authors used the average of three absolute standardized mean differences (ASMD) to assess the covariate balance of triplets after matching. Similarly, we can employ the average of $\text{ASMD} (\Xv^\ast, \Xv_1^\ast)$, $\text{ASMD} (\Xv^\ast, \Xv_2^\ast)$, and $\text{ASMD} (\Xv_1^\ast, \Xv_2^\ast)$ to evaluate the covariate balance of our matching.

Furthermore, we need to check the degree to which the distribution of observed pre-exposure covariates is similar across all exposure levels \citep{wu2022matching}. Thus, we divide the entire exposure range into $K$ blocks and calculate the block absolute standardized mean difference (BASMD) for each block. By aggregating the BASMD values for each exposure block, the average BASMD provides a comprehensive assessment of covariate balance across the entire dataset.

To be specific, the exposure range $\Z$ is categorized into $K$ blocks, denoted by $\Z_k$, where $k = 1,2,\dots,K$. The choice of how to block the exposure range can be based on different criteria. For example, the exposure range can be divided into blocks with the same number of samples in each block, or blocks with the same width. The BASMD can be defined as the ASMD of each blocks, $\text{BASMD}_k (\Xv_1^\ast, \Xv_2^\ast) = \frac{\vert\Bar{\Xv}_{k_1}-\Bar{\Xv}_{k_2}\vert}{\sqrt{(S_{k_1}^2+S_{k_2}^2)/2}}$ where $\Bar{\Xv}_{k_j}$ and $S_{k_j}^2$ denote the mean and sample variance of the covariates $\Xv_{i_j}$, respectively, for all $i$ such that $Z_i \in \Z_k$ and $j = 1,2$. $\text{BASMD}_k (\Xv^\ast, \Xv_1^\ast)$ and $\text{BASMD}_k (\Xv^\ast, \Xv_2^\ast)$ can be similarly defined. To measure the overall covariate balance of a triplet, we define $\text{BASMD}_k (\Xv^\ast, \Xv_1^\ast, \Xv_2^\ast) = \{\text{BASMD}_k (\Xv^\ast, \Xv_1^\ast) + \text{BASMD}_k (\Xv^\ast, \Xv_2^\ast) + \text{BASMD}_k (\Xv_1^\ast, \Xv_2^\ast)\}/3$. Then, we can use a criterion of the mean of the BASMD for each covariate being less than 0.1 as a threshold of good covariate balance across the whole exposure levels.
\begin{align*}
    \frac{1}{K}\sum_{k=1}^K \abs{\text{BASMD}_k (\Xv^\ast, \Xv_1^\ast, \Xv_2^\ast)} \leq (0.1,0.1,\dots,0.1)'.
\end{align*}

This is similar to the common criterion in the literature of having the standardized mean difference less than 0.1 when the exposure is binary. By averaging the BASMD of each covariate, we obtain the average BASMD as
\begin{align}\label{eqn:averageBASMD}
    \text{Average BASMD} = \overline{\frac{1}{K}\sum_{k=1}^K \abs{\text{BASMD}_k (\Xv^\ast, \Xv_1^\ast, \Xv_2^\ast)}}
\end{align}
where $\Bar{\Av}$ denotes the mean of all elements of the vector $\Av$. This metric can be used to assess the balance of covariates in the matched triplets, ensuring covariate balance. Consequently, we can select hyperparameters that minimize the average BASMD. In summary, we propose the matching algorithm for estimating the ACDE, as outlined in Algorithm \ref{alg1}.
    
\begin{algorithm}
    \caption{A Matching Algorithm for Average Causal Derivative Effect}\label{alg1}
    \begin{algorithmic}
        \setstretch{1.2}
        \renewcommand{\algorithmicrequire}{\textbf{Input:}}
	\renewcommand{\algorithmicensure}{\textbf{Output:}}
        \Require Data $\mathcal{D} = \{(\xv_i, y_i, z_i)\}_{i=1}^N$, a metric $d(\cdot, \cdot)$, an exposure level of interest $z_0\in\Z$, and blocks of $Z$: $Z_k$, $k = 1,\dots,K$.
        \Ensure Matched set $\{(i_1, i_2)\}_{i=1}^N$ and an estimated causal derivative effect at $z_0$, $\hat\delta(z_0)$.
        \State (a) Specify a candidate set of $(\eta,\kappa)$ with small $\eta(>0)$ values and $\kappa$ ranging from 0 to 1, or fix $\kappa$ to 0.1 by the rule of thumb.
        \State (b) For each pair of $(\eta, \kappa)$ from the candidate set, 
        \For{$i = 1,2,\dots, N$}
            \State Select $i_1$ and $i_2$ from the set $\{1, 2, \dots, N\}$ satisfying
        \begin{align*}
            i_1 = \argmin_{l:z_0-\eta\leq Z_l\leq z_0-\kappa\eta} d(\Xv_l, \Xv_i) \quad\text{and}\quad
            i_2 = \argmin_{l:z_0+\kappa\eta\leq Z_l\leq z_0+\eta} d(\Xv_l, \Xv_i)
        \end{align*}
        \EndFor
        \State (c) Calculate the average BASMD by \eqref{eqn:averageBASMD} for each matched dataset from step (b).
        \State (d) Choose the optimal $(\eta, \kappa)$ that minimizes the average BASMD and determine the corresponding matched set $\{(i_1, i_2)\}_{i=1}^N$, ensuring the best covariate balance.
        \State (e) Using the matched set $\{(i_1, i_2)\}_{i=1}^N$ in (d), estimate the ACDE at $z_0$ as $\hat\delta(z_0) = \frac{1}{N}\sumi\frac{Y_{i_2}-Y_{i_1}}{Z_{i_2}-Z_{i_1}}$.
    \end{algorithmic}
\end{algorithm}

\section{Permutation Test and Sensitivity Analysis for Local Causal Effect}\label{sec4}
To determine the significance of an exposure effect at a specified interest level of $Z = z_0$, we suggest a simple permutation-based testing procedure. The null hypothesis is that the exposure has no local effect on the outcome of interest, expressed as $H_0:Y_i(z_0)=Y_i(z_0+h), \forall i, \abs{h}\leq\eta$. We estimated the ACDE as a weighted mean of the average rates of change for all possible combinations of individuals with $Z_i$ in the intervals $[z_0 - \eta, z_0 - \kappa\eta]$ and $[z_0 + \kappa\eta, z_0 + \eta]$. To be specific, we introduce some notations. $\mathcal{S}_1$ represents the set of individuals matched on the left side of the exposure level $z_0$, while $\mathcal{S}_2$ is similarly defined for the right side. The set $\mathcal{S}$ encompasses every possible combination of matches from $\mathcal{S}_1$ and $\mathcal{S}_2$. The weight assigned to a matched combination $(k,l)$ is denoted as $w_{k,l}$, and $\delta_{k,l}$ signifies the average rate of change between individual $k$ and $l$.

\begin{gather*}
    \mathcal{S}_1 = \{1_1, \dots, N_1\}, \quad\mathcal{S}_2 = \{1_2, \dots, N_2\},\\
    \mathcal{S} = \{(k, l) : k\in \mathcal{S}_1, l \in \mathcal{S}_2\},\\
    w_{k,l} = \frac{1}{N}\sum_{i = 1}^N \Ind(i_1 = k)\cdot \Ind(i_2 = l),\\
    \delta_{k,l} = \frac{Y_k - Y_l}{Z_k - Z_l}.
\end{gather*}

Then, the ACDE can be estimated as:
\begin{align}\label{eqn:ACDE3}
    \hat\delta(z_0) = \sum_{(k,l)\in\mathcal{S}}w_{k,l}\cdot\delta_{k,l}.
\end{align}

\subsection{Permutation Test}\label{sec4.1}\hfill\break
This subsection introduces the ACDE as the test statistic for the permutation test, $T_N = \hat\delta(z_0)$. There are a total of $\abs{\Scal}$ combinations, each comprising two matched individuals, denoted as $k\in\Scal_1$ and $l\in\Scal_2$. The fundamental idea is to simply permute the sign of $\delta_{k,l}$ while preserving the matching structure. The set $\Fcal_{k,l} = \{(Y_j(z), \Xv_j, \Uv_j): z\in\Z , j \in \{k,l\}\}$ contains the relevant information for the individuals in the study, including potential outcomes $Y_j(z)$ at each value of $z\in\Z$, observed covariates $\Xv_j$, and any possible unobserved confounders $\Uv_j$ for each individual $j\in\{k,l\}$. The set $\Zcal_{k,l}$ includes two possible values of $\zv_{k,l} = (z_k, z_l)'$ of $\Zv_{k,l}$, such that $\{z_k, z_l\} = \{Z_k, Z_l\}$. If there is no hidden confounder, the exposure assignment $\Zv_{k,l}$ is selected randomly from $\Zcal_{k,l}$ with a probability of 1/2. Here, let $T_N = \hat\delta(z_0) = G(\Zv, \Yv)$ be a function of the vector of treatment assignments $\Zv$ and the vector of observed responses $\Yv$. The test statistic $T_N$ can be interpreted as an observed sample derived from a randomization distribution, which can be computed through permuted samples of $\Zv_{k,l}$. Alternatively, we can employ the normal approximation, by using the following theorem derived from the Hajek-Sidak Central Limit Theorem.
\begin{theorem}\label{thm2}
    If there is no local causal effect at $z_0$, 
    \begin{align*}
        \left(\sum_{(k, l)\in\Scal} w_{k,l}^2\cdot \delta_{k,l}^2\right)^{-1/2} T_N \xrightarrow{d} \Ncal(0,1)
    \end{align*}
    holds if $\frac{\max w_{k,l}^2\delta_{k,l}^2}{\sum w_{k,l}^2\delta_{k,l}^2}  \xrightarrow{} 0$ as $N$ goes infinity. 
\end{theorem}

If we restrict the maximum allowable matching number or ensure that the density function of $\Xv$ is bounded and bounded away from 0, we can attain the fulfillment of the ``if" condition by controlling the weight term. According to Theorem \ref{thm2}, we can compute the p-value to test the null hypothesis of no local causal effect at $z_0$. A practical approach is to select a value of $z_0$ based on the exposure variable, where the p-value falls below a pre-defined threshold, such as 0.05. This threshold serves as a criterion for determining the significance of the local causal effect.

\subsection{Sensitivity Analysis}\label{sec4.2}\hfill\break
In this section, we propose a method for conducting sensitivity analysis with continuous exposures, aimed at quantifying the impact of unmeasured confounders. The generalized propensity score (GPS), as introduced by \cite{hirano2004propensity}, serves as a tool for this sensitivity analysis. The GPS is defined as the conditional probability $e(z, \xv) = f_{Z_i\vert\Xv_i}(z\vert\xv)$, where $z \in \Z$ and $\xv \in \X$. Our method does not require knowledge of the exact GPS or its underlying model. Instead, it focuses on bounding the difference of the GPS within each matched combination.

If there is no unmeasured confounder, then $e(z,\xv_i) = e(z,\xv_j)$ holds for all $z\in\Z$ whenever $\xv_{i} = \xv_{j}$. However, in reality, observing all relevant covariates is challenging due to the complexity of the underlying framework and the presence of unmeasured confounders. To account for this, we introduce a sensitivity parameter $\Gamma$, which bounds the potential deviation from this assumption arising from unmeasured confounders. Specifically, whenever $\xv_{i} = \xv_{j}$, the following inequality holds:
\begin{align}\label{eqn:sensmodel}
\frac{1}{\Gamma} \leq \frac{e(z',\xv_{i})\cdot e(z,\xv_{j})}{e(z,\xv_{i})\cdot e(z',\xv_{j})} \leq \Gamma, \quad \forall z, z'\in [z_0 - \eta, z_0 + \eta].
\end{align}

This model serves as an extension of the sensitivity model proposed by \cite{rosenbaum1987sensitivity} for binary exposures, adapting it to handle continuous exposures. The sensitivity parameter $\Gamma$ provides a quantification of the potential impact of unmeasured confounders on the observed generalized propensity scores within a specified neighborhood around $z_0$. Since 
\begin{align}\label{eqn:sensprob}
   \Prob(Z_{k}=z_{k}\vert\Fcal_{k,l},\Zcal_{k,l}) = \frac{e(z_{k},\xv_{k})e(z_{l},\xv_{l})}{e(z_{k},\xv_{k})e(z_{l},\xv_{l}) + e(z_{l},\xv_{k})e(z_{k},\xv_{l})} 
\end{align}
holds, \eqref{eqn:sensmodel} implies that
\begin{align}\label{eqn:sensprobbound}
\frac{1}{1+\Gamma} \leq \Prob(Z_{k}=z_{k}\vert\Fcal_{k,l},\Zcal_{k,l}) \leq \frac{\Gamma}{1+\Gamma}.
\end{align}

Although the exact probability $p_{k,l} = \Prob(Z_{k}=z_{k}\vert\Fcal_{k,l},\Zcal_{k,l})$ may be unknown, it can be bounded by \eqref{eqn:sensprobbound} with the sensitivity parameter $\Gamma$. Define $p_{k,l}^+ = \frac{\Gamma}{1+\Gamma}$ and $p_{k,l}^- = \frac{1}{1+\Gamma}$, satisfying $p_{k,l}^- \leq p_{k,l} \leq p_{k,l}^+$ for $n = 1,2,\dots, N$. Define $T^+$ as the sum of $\abs{\Scal}$ independent random variables, where each $(k,l)$th term takes the value $w_{k,l}\cdot\abs{\delta_{k,l}}$ with probability $p_{k,l}^+$ and $-w_{k,l}\cdot\abs{\delta_{k,l}}$ with probability $1-p_{k,l}^+$. Define $T^-$ similarly with $p_n^-$ in place of $p_n^+$.

\begin{theorem}\label{thm3}
If there is no local causal effect at $z_0$, then for each fixed $\Gamma\geq 1$,
\begin{align*}
\Prob(T^+\geq a) \geq \Prob(T_N\geq a\vert \Fcal, \Zcal)\geq \Prob(T^-\geq a).
\end{align*}
\end{theorem}

This can be seen as a generalization of a permutation test introduced in Section~\ref{sec4.1}, where we set $\gamma = 1$. The precise distributions of $T^+$ and $T^-$ can be established by computing the randomization distribution through permutation or by employing a normal approximation, similar to the approach described in the previous section. Therefore, we can compute the upper bound on the p-value for each $\Gamma$ based on permuted samples. However, due to computational limitations, we may use an asymptotic approximation to establish upper and lower bounds on $\Prob(T_N \geq a\vert \Fcal, \Zcal)$ based upon a normal approximation, similar to Theorem~\ref{thm2}. Depending on this value, we can account for the effect of unmeasured confounder while estimating the derivative effect. This sensitivity analysis method provides a manner to measure the potential impact of unmeasured confounders on the estimated derivative effect, allowing researchers to quantify the degree to which unmeasured confounding may affect their results.

\subsection{Simulation Studies: Permutation Test}\label{sec4.3}\hfill\break
In this subsection, we conduct simulation studies to investigate the performance of proposed permutation test and sensitivity analysis methods. We apply proposed methods to investigate the practical applicability of our approach in estimating the ACDE and determining a threshold for the exposure level that significantly influences the outcome of interest. The data generation process involves three independent covariates, denoted as $(X_{i1}, X_{i2}, X_{i3})$, each drawn from independent standard normal distributions. The exposure and outcome are then generated using the following equations:
\begin{align}\label{eqn:sim3}
Z_i &= 5 + (X_{i1} + X_{i2} + X_{i3}) + \varepsilon_{z,i},\\\nonumber
Y_i &= 3 + 15 (X_{i1} - X_{i2} + X_{i3}) + (Z_i - 5)^2 \cdot \mathbb{I}(Z_i-5>0) + \varepsilon_{y,i},\\\nonumber
\varepsilon_{z,i} &\sim \mathcal{N}(0, 4^2), \quad \varepsilon_{y,i} \sim \mathcal{N}(0, 1^2),
\end{align}
where $\varepsilon_{z,i}$ and $\varepsilon_{y,i}$ represent errors generated from normal distributions. The average causal derivative effect is calculated as $\delta(z_0) = 2(z_0-5) \cdot \mathbb{I}(z_0-5>0)$, with $z_0$ representing the exposure level of interest.

\begin{figure}
\begin{center}
\includegraphics[width=3in]{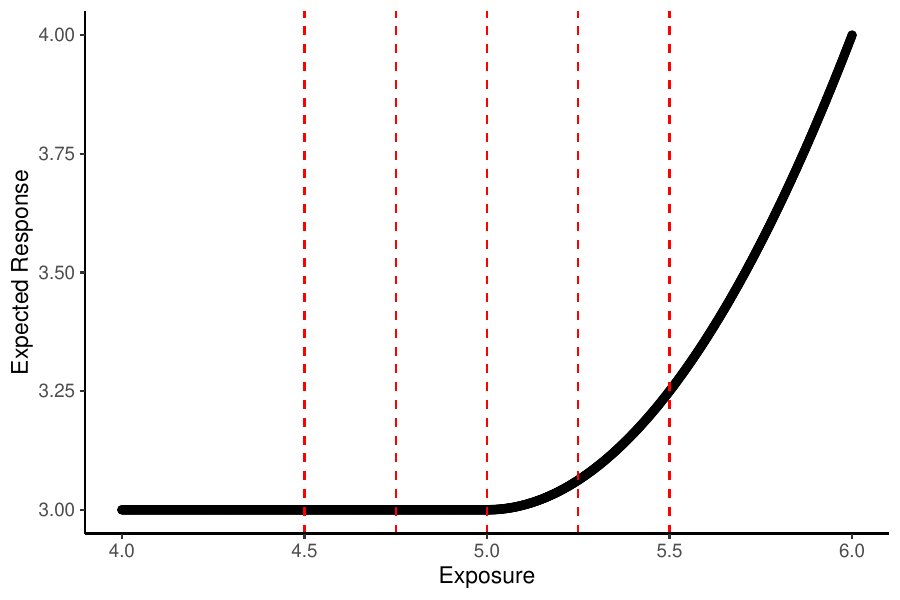}
\end{center}
\caption{Exposure-response function in the simulation settings. Target exposure levels are 4.5, 4.75, 5, 5.25, and 5.5.}\label{fig1}
\end{figure}

In this simulated scenario, our objective is to investigate the applicability of the proposed method in estimating the ACDE and determining whether it can effectively identify the exposure level that significantly influences an outcome. We know the true exposure-response curve as Figure~\ref{fig1}, that is, $\E[Y_i(z)] = (z-5)^2\cdot \mathbb{I}(z-5>0)$. Consequently, we can use an exposure level of 5 as a threshold to define a treated group. 

We examine the performance of our methods on exposure levels $z_0 =$ 4.5, 4.75, 5, 5.25, and 5.5, where the true population ACDEs are 0, 0, 0, 0.5, and 1, respectively. The parameters are set as follows: sample size $N = 3000$, $\eta = 0.5$, $\kappa = 0.1$, and covariate distance is measured using scaled Euclidean distance. In this setting, we skip the process of determining hyperparameters using the covariate balance approach. The evaluation is based on 500 simulations, and we report the absolute bias, mean square error (MSE), and the rejection rate of the hypothesis test with a significance level of 0.05 in Table~\ref{table1}.

At exposure levels other than 5, the ACDEs are estimated with relatively small bias and RMSE. Estimating the ACDE at 4.75 and 5.25 can be challenging since $\eta=0.5$ implies that matched individuals may exhibit bias, meaning individuals on the opposite side of 5 could be matched. However, the overall RMSE does not show significant differences at 4.75 and 5.25 compared to 4.5 and 5.5 even though the bias is slightly larger at 4.75 and 5.25, which implies that the ACDEs are well-estimated. However, the absolute bias reaches their maximum with high RSME at the exposure level of 5. This occurs because, even though the derivative effect at 5 is equal to 0, the average rate of change near 5 is not equal to 0. This issue can be mitigated as the sample size $N$ approaches infinity and the hyperparameter $\eta$ approaches toward 0. This property will be further discussed in Section~\ref{sec5}.

\begin{table}
	\centering
	\caption{Performance of the matching method in estimating the ACDE. We examine five exposure levels: 4.5, 4.75, 5, 5.25, and 5.5. The true ACDE, absolute bias, root mean square error (RMSE), and rejection rates ($\alpha=0.05$) are reported.}
	\begin{tabular}{c|ccccc}
		\hline
		& \multicolumn{5}{c}{Exposure Level} \\[0.1cm]
		& 4.5 & 4.75 & 5 & 5.25 & 5.5 \\[0.1cm]
		\hline
            True ACDE & 0 & 0 & 0 & 0.5 & 1\\[0.1cm]
            Absolute Bias & 0.142 & 0.182 & 0.416 & 0.208 & 0.156 \\[0.1cm]
            Root Mean Square Error & 1.758 & 1.595 & 1.794 & 1.798 & 1.886 \\[0.1cm]
            Rejection Rate & 0.064 & 0.050 & 0.088 & 0.100 & 0.148 \\[0.1cm]
		\hline
	\end{tabular}
	\label{table1}
\end{table}

\section{Asymptotic Properties of the Matching Estimator}\label{sec5}
\subsection{Asymptotic Properties}\label{sec5.1}\hfill\break
In this section, we explore the asymptotic properties of the proposed matching estimator, $\hat{\delta}_{z_0}$. We focus on a difference between $\hat\delta(z_0)$ and $\delta(z_0)$ for a fixed value of $z_0$ that we are interested in. Define the sample ACDE for finite sample $\{(Y_i,Z_i,\Xv_i)\}_{i=1}^N$ as
\begin{align}\label{eqn:sampleACDE}
    \Bar{\delta}(z_0) = \frac{1}{N}\sumi\E[Y_i'(z_0)\vert \Xv_i] = \frac{1}{N}\sumi\mu'_{\Xv_i}(z_0)
\end{align}
where the expectation is taken on potential outcomes conditional on $\Xv_i$. As the sample size $N$ goes to infinity, the sample ACDE $\Bar{\delta}(z_0)$ converges to the population ACDE $\delta(z_0)$. 

In \cite{abadie2006large}, the authors decomposed the difference between the true population ATE and its matching estimator in the context of binary exposure. Additionally, the difference between the population average causal exposure-response function and its matching estimator was decomposed by \cite{wu2022matching}. Similarly, the difference between $\hat\delta(z_0)$ and $\delta(z_0)$ can also be decomposed as follows:
\begin{align}\label{eqn:decomposition}
    \hat\delta(z_0)-\delta(z_0) = \{\Bar{\delta}(z_0)-\delta(z_0)\}+\Ecal_{z_0}+\Acal_{z_0}+\Bcal_{z_0},
\end{align}
where $\Ecal_{z_0}$ denotes a weighted average of the residuals,
\begin{align*}
    \Ecal_{z_0}=\frac{1}{N}\sumi\frac{\eps_{i_2}-\eps_{i_1}}{Z_{i_2}-Z_{i_1}},
\end{align*}
$\Acal_{z_0}$ denotes an error generated by an approximation,
\begin{align*}
    \Acal_{z_0} = \frac{1}{N}\sumi\{\frac{\mu_{\Xv_i}(Z_{i_2})-\mu_{\Xv_i}(Z_{i_1})}{Z_{i_2}-Z_{i_1}} - \mu'_{\Xv_i}(z_0)\},
\end{align*}
and $\Bcal_{z_0}$ is the bias generated by matching,
\begin{align*}
    \Bcal_{z_0}=\frac{1}{N}\sumi\{\frac{\mu_{\Xv_{i_2}}(Z_{i_2})-\mu_{\Xv_i}(Z_{i_2})}{Z_{i_2}-Z_{i_1}} - \frac{\mu_{\Xv_{i_1}}(Z_{i_1})-\mu_{\Xv_i}(Z_{i_1})}{Z_{i_2}-Z_{i_1}}\}.
\end{align*}

The overall goal for this section is to prove the consistency and asymptotic normality of our matching estimator in some specific instances. We use the Euclidean distance $d(\cdot, \cdot)$ while fixing the caliper size $\kappa$ as a constant with respect to $N$ and $\eta = o(N^{-\frac{1}{d+3}})$, allowing us to show that the proposed matching estimator is consistent. If there is only one covariate being matched, the estimator is also asymptotically normal and converges at a rate of $(N\eta^3)^{-1/2}$. However, when more than one covariate is matched, an approximation error and bias must be removed to achieve asymptotic normality. This study builds upon previous works by \cite{abadie2006large, wu2022matching}, but focuses on the matching estimator for the ACDE with continuous exposures. Additional regularity assumptions necessary to ensure asymptotic properties and detailed discussions are addressed in the supplementary materials.

\begin{theorem}[Consistency]\label{thm4}
    Assume Assumptions~\ref{ass1}--\ref{ass4} and S.1 in the supplementary materials hold. Then, $\hat\delta(z_0)-\delta(z_0) \xrightarrow{p} 0$ holds.
\end{theorem}

Theorem~\ref{thm4} shows that the matching estimator $\hat\delta(z_0)$ is consistent. Note that we can consistently estimate the true population ACDE regardless of the dimension of covariates.

\begin{theorem}[Asymptotic Normality]\label{thm5}
    Assume Assumptions~\ref{ass1}--\ref{ass4} and S.1 in the supplementary materials hold. Then, the following holds.
    \begin{gather*}
        \Sigma_{z_0}^{-1/2}(N\eta^3)^{1/2}\{\hat\delta(z_0)-\Bcal_{z_0}-\Acal_{z_0}-\delta(z_0)\} \xrightarrow{d} \Ncal(0,1), \\
        \Sigma_{z_0} = \frac{1}{N}\sumj \eta^3 I_j  L_{z_0}(j)^2 \sigma^2(Z_j, \Xv_j) 
    \end{gather*}
\end{theorem}
where $\Lj$ is the sum of all coefficients of $Y_j$ in $\sumi\hat\delta_i(z_0)$ and $I_j$ is an indicator of whether an individual $j$ can be used for matching while estimating $\delta(z_0)$ (see supplementary materials for detailed notations). Even though the proposed matching estimator is consistent, Theorem~\ref{thm5} states that in order to guarantee asymptotic normality, it is necessary to remove the conditional bias and approximation term. This method can be accomplished through parametric or nonparametric estimations of $\mu(z,\xv)$, similar to the bias-corrected matching estimators for average treatment effects \citep{abadie2011bias}. Additionally, this theorem suggests that when matching scalar covariates, the bias and approximation terms can be neglected.

\begin{corollary}[Asymptotic Normality with Scalar Covariates]\label{cor1}
    Assume Assumptions~\ref{ass1}--\ref{ass4} and S.1 in the supplementary materials hold. If $\Xv_i$ is scalar, the following holds.
    \begin{gather*}
        \Sigma_{z_0}^{-1/2}(N\eta^3)^{1/2}\{\hat\delta(z_0)-\delta(z_0)\} \xrightarrow{d} \Ncal(0,1).
    \end{gather*}
\end{corollary}

Corollary~\ref{cor1} implies that when the covariate consists of only one continuous variable, the matching estimator is $(N\eta^3)^{1/2}$-consistent and asymptotically normal. Reducing the dimension of covariates results in a faster convergence rate of the estimator and even leads to asymptotic normality in the scalar covariate setting. A possible extension is to use generalized propensity score matching, which has the advantage of reducing the dimension of matching \citep{abadie2016matching}.

\subsection{Simulation Studies: Asymptotic Properties}\label{sec5.2}\hfill\break
In this subsection, we expand the simulation studies presented in Section~\ref{sec4.3} to explore the performance of the proposed methods under varying sample sizes and covariate dimensions. We examine three distinct sample sizes: $N = 3000, 6000, 10000$, and three categories of covariate dimensions: $d = 2, 3, 4$. We focus on the exposure level of 5, which is our exposure level of interest. For the model of dimension 3, we use the model \eqref{eqn:sim3} again. Similarly, we employ similar models:
\begin{align*}
Z_i &= 5 + (X_{i1} + X_{i2}) + \varepsilon_{z,i},\\
Y_i &= 3 + 15 (X_{i1} - X_{i2}) + (Z_i - 5)^2 \cdot \mathbb{I}(Z_i-5>0) + \varepsilon_{y,i}
\end{align*}
for the covariate dimension of 2 cases, and
\begin{align*}
Z_i &= 5 + (X_{i1} + X_{i2} + X_{i3} -0.5 X_{i4}) + \varepsilon_{z,i},\\
Y_i &= 3 + 15 (X_{i1} - X_{i2} + 2 X_{i3} + 2X_{i4} ) + (Z_i - 5)^2 \cdot \mathbb{I}(Z_i-5>0) + \varepsilon_{y,i}
\end{align*}
for the covariate dimension of 4 cases. The covariates are randomly generated from the standard normal distribution, and error terms are the same with the model \eqref{eqn:sim3}. We use values of $\eta$ set at 0.5, 0.45, and 0.4 for each $N$ of 3000, 6000, and 10000, respectively, while keeping $\kappa$ fixed at 0.1. The results, including absolute bias, root mean squared error (RMSE), and rejection rates, are presented in Table~\ref{table2}.
\begin{table}
	\centering
	\caption{Performance of the matching method in estimating the ACDE at $z_0 = 5$. We examine three sample sizes: 3000, 6000, 15000, and three covariate dimensions: 2, 3, 4. The absolute bias, root mean square error (RMSE), and rejection rates ($\alpha=0.05$) are reported.}
	\begin{tabular}{ccccccc}
            \hline
            $N$ & $\eta$ & $\kappa$ & & \multicolumn{3}{c}{Covariate Dimension}\\[0.1cm]
		& & & & 2 & 3 & 4\\[0.1cm]
		\hline
            3000 & 0.5 & 0.1 & Absolute Bias & 0.187 & 0.416 & 0.395\\[0.1cm]
             & & & RMSE & 0.720 & 1.794 & 3.150\\[0.1cm]
             & & & Rejection Rate & 0.046 & 0.088 & 0.130\\[0.1cm]
             \hline
            6000 & 0.45 & 0.1 & Absolute Bias & 0.114 & 0.290 & 0.182\\[0.1cm]
             & & & RMSE & 0.473 & 1.257 & 2.189\\[0.1cm]
             & & & Rejection Rate & 0.062 & 0.070 & 0.126\\[0.1cm]
             \hline
            10000 & 0.4 & 0.1 & Absolute Bias & 0.115 & 0.224 & 0.294\\[0.1cm]
             & & & RMSE & 0.403 & 0.903 & 1.660\\[0.1cm]
             & & & Rejection Rate & 0.060 & 0.058 & 0.082 \\[0.1cm]
		\hline
	\end{tabular}
	\label{table2}
\end{table}

As $N$ increases and the value of $\eta$ decreases, the proposed matching estimator converges to the true value. Additionally, the rejection rates approach the desired level of 0.05. However, as highlighted in this section, the dimension plays a crucial role. With an increase in the covariate dimension, both the computational cost of the matching method and the root mean squared error (RMSE) rise. The rate of convergence is fast in the low-dimensional case. Nevertheless, irrespective of the covariate dimension, the method remains consistent, and the test procedure is asymptotically valid as the sample size increases. 

\section{Data Application}\label{sec6}
Chronic Obstructive Pulmonary Disease (COPD) represents a significant global health burden, characterized by persistent and progressive airway inflammation \citep{im2023causal}. The primary pathological features of COPD include emphysema and small airway obstruction and destruction, both quantifiable through CT-assessed emphysema and functional small airway disease (fSAD) utilizing parametric response mapping (PRM). \cite{im2023causal} conducted a causal inference analysis on longitudinal data to investigate the progression of the disease, specifically focusing on the causal effect of fSAD in the advancement of emphysema. The study group was stratified into low and high total percent fSAD, employing a baseline cut-off value of 15\%. However, the selection of the cut-off value at 15\% appears arbitrary and lacks sufficient support from previous literature. Consequently, we will employ our matching method for estimating the ACDE to identify the optimal fSAD level for stratification into low and high fSAD groups. This approach aims to provide a more data-driven and meaningful threshold for distinguishing between the two groups based on their causal impact on disease progression.

In this section, we apply our causal framework of the ACDE to the data set of 128 male COPD patients. We use the same data of \cite{im2023causal}, which was derived from independent cohorts from university-affiliated hospitals in Seoul, Republic of Korea. A derivation cohort was constructed based on a retrospective single-center cohort of patients with COPD from January 2016 to April 2020 at Samsung Medical Center in Seoul, Republic of Korea. COPD was defined as post-bronchodilator FEV1/forced vital capacity of less than 70\%, without current asthma. Independent data were obtained from the Korean Obstructive Lung Disease cohort, a multicenter cohort from 16 hospitals in the Republic of Korea from June 2005 to October 2015. In this dataset, we measure the total percentages of fSAD and emphysema twice for each individual. The baseline total percent fSAD serves as the exposure variable, and the difference in total emphysema is considered as the outcome of interest. The analysis includes three covariates: age, BMI, and smoking status. Here, age and BMI are continuous covariates, while smoking status is binary (1 if yes, 0 if no).

We calculate the ACDE at various levels of fSAD, ranging from 0.04 to 0.2. For each fSAD level, we determine optimal hyperparameters, $\eta$ and $\kappa$, based on their average BASMDs. The hyperparameters, $\eta$ and $\kappa$, are selected from the grid $\eta \in \{0.04, 0.045, 0.5, 0.055, 0.6, 0.65, 0.7, 0.75, 0.8\}$ and $\kappa \in \{0.1, 0.15, 0.2\}$. To measure an average BASMD, the dataset is divided into four blocks, $Z_1, Z_2, Z_3, Z_4$, each containing 32 individuals, based on quantiles of fSAD levels. Using these optimized hyperparameters, we estimate the ACDEs, and their associated p-values are calculated through a permutation test we proposed. The summarized results are presented in Table~\ref{table3}.

\begin{figure}
    \begin{center}
        \includegraphics[width=5.5in]{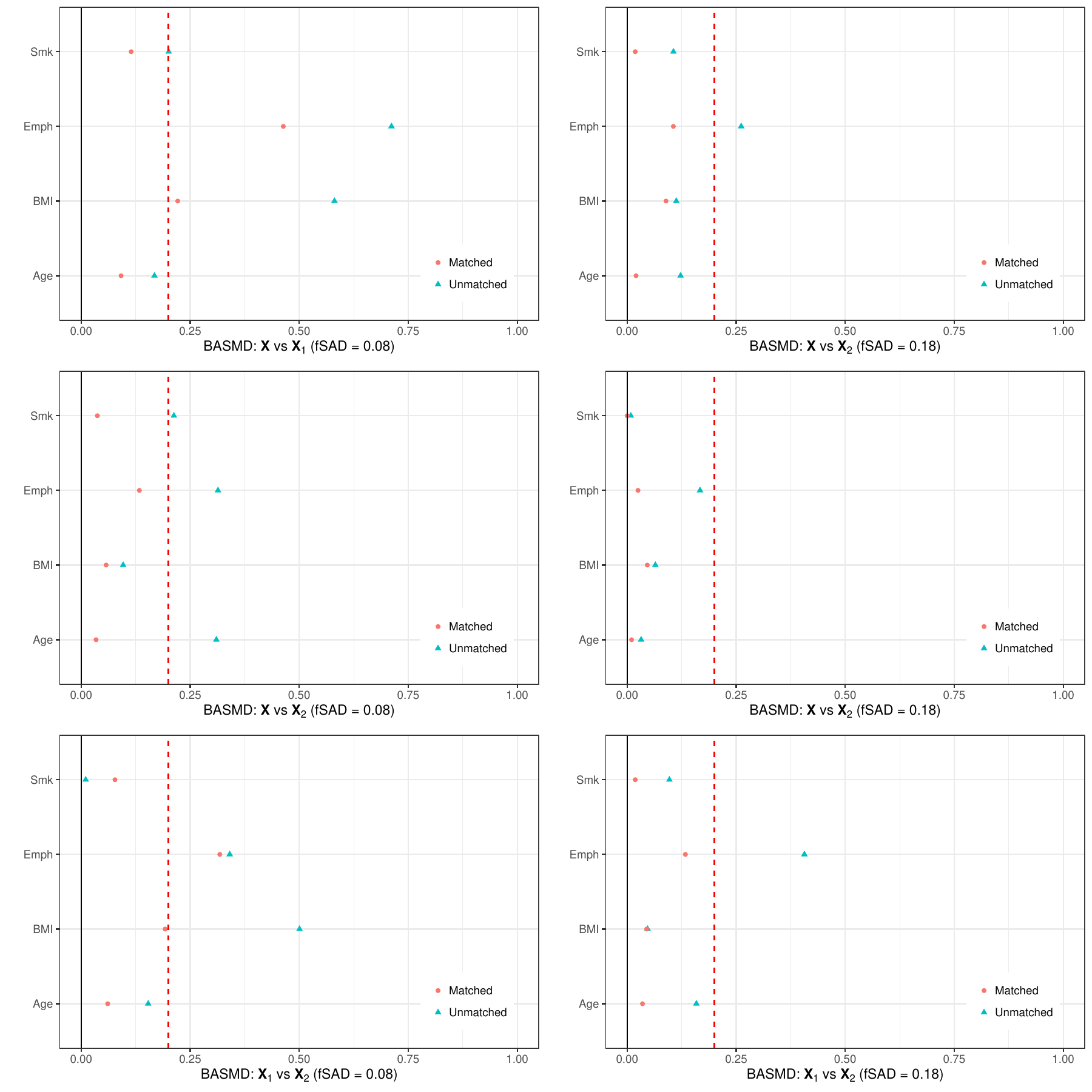}
    \end{center}
\caption{Average BASMDs for (1) $\Xv$ vs $\Xv_1$, (2) $\Xv$ vs $\Xv_2$, and (3) $\Xv_1$ vs $\Xv_2$ at \\fSAD = 0.08 (left) and fSAD = 0.18 (right) with optimal hyperparameters.}\label{fig2}
\end{figure}

The ACDE reaches its maximum significance at an fSAD level of 0.08 (ACDE = $0.285$, p-value = $0.001$). In comparison to \cite{im2023causal}, where a dichotomized fSAD level with a threshold of 0.15 was employed, our results indicate that even within the proximity of the fSAD level of 0.08, a significant impact on the progression of emphysema is observed. Therefore, we propose considering a threshold of 0.08 to categorize COPD patients into high/low fSAD groups. Notably, as the fSAD level continues to increase, the ACDE becomes negative. Especially the ACDE shows the most significant negative effect at the fSAD level of 0.18 (ACDE = $-0.288$, p-value = $0.017$). Among these levels, we specifically focus on fSAD levels of 0.08 and 0.18 with the optimal hyperparameters. The BASMDs for (1) $\Xv$ vs $\Xv_1$ (Original group vs $\mathcal{S}_1$ group), (2) $\Xv$ vs $\Xv_2$ (Original group vs $\mathcal{S}_2$ group), and (3) $\Xv_1$ vs $\Xv_2$ ($\mathcal{S}_1$ group vs $\mathcal{S}_2$ group) are illustrated in Figure~\ref{fig2} for both pre-matched and matched sets. At the fSAD level of 0.18, the covariates in the three groups are balanced well. However, at the fSAD level of 0.08, achieving covariate balance is challenging. In this case, we may apply covariate adjustment before the analysis \citep{rosenbaum2002covariance} to reduce a potential bias arising from this covariate imbalance. Detailed discussions and the application of the covariate adjustment method are provided in the supplementary materials.

\begin{table}
	\centering
	\caption{Estimation of the ACDE of emphysema is conducted at each level of fSAD. The optimal values for $\eta$ and $\kappa$, determined based on the average BASMD, are reported. Additionally, p-values are calculated using the permutation test.}
	\begin{tabular}{cccccc}
		\hline
		fSAD & $\eta$ & $\kappa$ & Average BASMD & ACDE & p-value\\[0.1cm]
		\hline
          0.04 & 0.07 & 0.2 & 0.413 & -0.072 & 0.669\\[0.1cm]
            \textbf{0.06} & 0.08 & 0.1 & 0.334 & \textbf{0.206} & \textbf{0.003}\\[0.1cm]
            \textbf{0.08} & 0.08 & 0.15 & 0.170 & \textbf{0.285} & \textbf{0.001}\\[0.1cm]
            0.10 & 0.08 & 0.15 & 0.172 & 0.063 & 0.236\\[0.1cm]
            0.12 & 0.075 & 0.15 & 0.122 & 0.010 & 0.893\\[0.1cm]
            0.14 & 0.075 & 0.1 & 0.138 & 0.012 & 0.918\\[0.1cm]
            0.16 & 0.055 & 0.1 & 0.140 & -0.197 & 0.346\\[0.1cm]
            \textbf{0.18} & 0.075 & 0.1 & 0.083 & \textbf{-0.288} & \textbf{0.017}\\[0.1cm]
            0.20 & 0.065 & 0.15 & 0.107 & -0.323 & 0.065\\[0.1cm]
            \hline
	\end{tabular}
	\label{table3}
\end{table}

\begin{table}
	\centering
	\caption{Sensitivity analysis for unmeasured confounders at fSAD levels of 0.08 and 0.18.}
	\begin{tabular}{c|cc|cc}
		\hline
               & \multicolumn{2}{c}{fSAD = 0.08} & \multicolumn{2}{c}{fSAD = 0.18}\\[0.1cm]
              \hline
		   $\Gamma$& Lower p-value & Upper p-value & Lower p-value & Upper p-value\\[0.1cm]
		\hline
            1.000 & \textbf{0.001} & \textbf{0.001} & \textbf{0.017} & \textbf{0.017}\\[0.1cm]
            1.050 & $<\textbf{0.001}$ & \textbf{0.003} & \textbf{0.009} & \textbf{0.028}\\[0.1cm]
            1.100 & $<\textbf{0.001}$ & \textbf{0.007} & \textbf{0.005} & \textbf{0.045}\\[0.1cm]
            1.110 & $<\textbf{0.001}$ & \textbf{0.008} & \textbf{0.004} & \textbf{0.050}\\[0.1cm]
            1.150 & $<\textbf{0.001}$ & \textbf{0.014} & \textbf{0.003} & 0.069\\[0.1cm]
            1.200 & $<\textbf{0.001}$ & \textbf{0.027} & \textbf{0.002} & 0.100\\[0.1cm]
            1.250 & $<\textbf{0.001}$ & \textbf{0.047} & $<\textbf{0.001}$ & 0.139\\[0.1cm]
            1.255 & $<\textbf{0.001}$ & \textbf{0.050} & $<\textbf{0.001}$ & 0.143\\[0.1cm]
            \hline
	\end{tabular}
	\label{table4}
\end{table}

We also conducted a sensitivity analysis as outlined in Section~\ref{sec4.2} at two fSAD levels, 0.08 and 0.18, to investigate the impact of unmeasured confounders. The summarized results are presented in Table~\ref{table4}. In summary, both points are statistically significant when no unmeasured confounders are considered. However, in the presence of unmeasured confounders, the statistical conclusions may vary. For instance, when $\Gamma = 1.11$, the upper p-value for fSAD = 0.18 is precisely 0.05. This indicates that if unmeasured confounders exist with an extent greater than $\Gamma = 1.11$, the ACDE no longer retains its significance at the fSAD level of 0.18. Similarly, when $\Gamma$ exceeds 1.255, the upper p-value for fSAD = 0.08 becomes less than 0.05.

\section{Conclusion}\label{sec7}
The estimation of causal effects is a fundamental challenge in various scenarios. When exposure variable is binary, causal effects are often explained through deterministically intervened estimands, such as the average treatment effect or the average treatment effect on the treated. However, there is a lack of prior research focused on the causal effects associated with exposures on a continuous scale. In this paper, we propose a deterministically intervened causal estimand, the average causal derivative effect (ACDE), and introduce a causal inference framework centered around this estimand. We present a matching method for estimating the ACDE, offering the advantage of extending this matching structure to include permutation tests for no local causal effects and conducting sensitivity analyses. To the best of our knowledge, there is limited research on testing and sensitivity analysis with continuous exposures in the field of causal inference. Finally, we demonstrate that the estimator is consistent and follows asymptotic normality under specific conditions. As an example, we apply this framework to data from COPD patients, providing insights into establishing a threshold for dichotomizing individuals into high and low exposure groups.

Our findings suggest that at the fSAD level of 0.08, fSAD has the most significant effect on the progression of emphysema. This implies potential policies focusing on COPD patients with an fSAD level of 0.08. Additionally, this value can serve as a threshold for categorizing COPD patients into high/low fSAD groups in some cases. Furthermore, the causal effect at this level remains significant until considering unmeasured confounders of $\Gamma = 1.255$. Given the limited number of causal inference approaches for fSAD and emphysema levels, this consideration can serve as a valuable guideline.

In contrast to the exposure-response function, our approach allows the direct estimation of the expected causal effect of an exposure on the outcome. Through a hypothesis testing procedure, we can validate the significance of the proposed estimator. However, some of the assumptions may be unrealistic in many real cases. For instance, the overlap assumption may fail if certain units are almost impossible to have an exposure level of interest. Additionally, if the exposure-response function is not smooth, the target estimand may not be well-defined. These issues can be addressed by targeting another estimand, such as the derivative effects of specific individuals, which is similar to the average treatment effect on the treated (ATT) in binary settings. We leave these potential improvements as subjects for future work.

\section*{Supporting Information}
The R-codes described in this paper are available at \url{https://github.com/suhwanbong121/average_causal_derivative_effect}. Supplementary materials, which include proofs, detailed discussions on asymptotic properties, extensions to 1:$M$ matching, and further insights into covariate adjustments, are available online.

\bibliographystyle{apalike}
\bibliography{references}

\pagestyle{empty}


\section*{Supplementary material}

# Supplementary Materials

## S.1 Assumptions and Proofs

### S.1.1 Regularity Assumptions

The following regularity assumption guarantees the asymptotic properties we desire.

**Assumption S.1** (Regularity Conditions)**.**

(a) Let $\mathbf{X}$ be a random vector of dimension d of continuous covariates distributed on $\mathbb{R}^d$ with compact and convex support $\mathbb{X}$, with density bounded and bounded away from zero on its support.

(b) $\mu(z, \mathbf{x})$ is Lipschitz-continuous in $\mathbf{x}$ and its second derivative with respect to $z$, $\frac{\partial^2}{\partial z^2}\mu(z, \mathbf{x})$, is uniformly bounded.

(c) $\underline{\sigma}^2(z, \mathbf{x}) \equiv \inf_{z,\mathbf{x}} \sigma^2(z, \mathbf{x}) > 0$, $\bar{\sigma}^2(z, \mathbf{x}) \equiv \sup_{z,\mathbf{x}} \sigma^2(z, \mathbf{x}) < \infty$.

(d) For some $\zeta > 0$, $\mathbb{E}\{|Y_i|^{2+\zeta} | Z_i = z, \mathbf{X}_i = \mathbf{x}\}$ is uniformly bounded.

### S.1.2 Proof of Theorem 1

*Proof.*

$$
\begin{aligned}
\frac{\partial}{\partial z}\mathbb{E}[Y_i(z)] &= \lim_{h \to 0} \frac{\mathbb{E}[Y_i(z+h)] - \mathbb{E}[Y_i(z)]}{h} \\
&= \lim_{h \to 0} \mathbb{E}\left[\frac{Y_i(z+h) - Y_i(z)}{h}\right] \\
&= \lim_{h \to 0} \mathbb{E}[Y_i'(z_h)]
\end{aligned}
$$

holds for some $z_h$ between $z$ and $z + h$. Since $Y_i'(z)$ is continuous and bounded, dominated convergence theorem implies

$$\lim_{h \to 0} \mathbb{E}[Y_i'(z_h)] = \mathbb{E}[\lim_{h \to 0} Y_i'(z_h)]$$

$$= \mathbb{E}[Y_i'(z)]$$

holds. $\qquad\qquad\qquad\qquad\qquad\qquad\qquad\qquad\qquad\qquad\qquad\qquad\qquad\qquad\square$

We also introduce some notations for algebraic simplicity. We define the conditional moments of potential outcomes given covariates and several functions as follows:

$$\mu_j = \mu(Z_j, \mathbf{X}_j), \quad \sigma_j^2 = \sigma^2(Z_j, \mathbf{X}_j)$$

$$K_{z_0}(j) = \sum_{k=1}^{N} \{\mathbb{I}(k_1 = j) + \mathbb{I}(k_2 = j)\}$$

$$L_{z_0}(j) = \sum_{k=1}^{N} \{\frac{\mathbb{I}(k_2 = j) - \mathbb{I}(k_1 = j)}{Z_{k_2} - Z_{k_1}}\}$$

$$I_j = \mathbb{I}(Z_j \in [z_0 - \eta, z_0 - \kappa\eta]) + \mathbb{I}(Z_j \in [z_0 + \kappa\eta, z_0 + \eta]).$$

where $K_{z_0}(j)$ is the number of times individual $j$ is used as a match, $L_{z_0}(j)$ is the sum of all coefficients of $Y_j$ in $\sum_{i=1}^{N} \hat{\delta}_i(z_0)$. Also, $I_j$ is an indicator of whether an individual $j$ can be used for matching while estimating $\delta(z_0)$. The matching estimator $\hat{\delta}(z_0)$ can then be represented as a weighted sum of $Y_j$,

$$\hat{\delta}(z_0) = \frac{1}{N} \sum_{j=1}^{N} L_{z_0}(j) Y_j I_j. \tag{1}$$

We begin by investigating the matching discrepancy. Abadie and Imbens (2006) showed that the discrepancy of the matching method is given as the following lemma.

**Lemma S.1** (Matching Discrepancy). *Let* $i_0 = \mathrm{argmin}_{1 \le i \le N} \|\mathbf{X}_i - \mathbf{x}\|$ *and define* $U =$

$\mathbf{X}_{i_0} - \mathbf{x}$ as a matching discrepancy. Assume Assumption S.1 holds, then all the moments of $N^{1/d} \|U\|$ are uniformly bounded in $N$ and $\mathbf{x} \in \mathbb{X}$.

*Proof.* This lemma follows from Abadie and Imbens (2006). $\qquad\square$

This lemma allows us to calculate the stochastic order of conditional bias, which is a key factor in determining the accuracy of the estimator.

**Lemma S.2** (The Order of Bias). Assume Assumptions 1–4 and S.1 hold. Then, the order of bias of the matching estimator is $\mathcal{B}_{z_0} = O_p\{(N\eta^{d+1})^{-1/d}\}$.

*Proof.* Since $\mu(z, \mathbf{x})$ is Lipschitz-continuous in $\mathbf{x}$,

$$\begin{aligned}
|\mathcal{B}_{z_0}| &\leq \frac{1}{N} \sum_{i=1}^{N} \left| \frac{\mu_{\mathbf{X}_{i_2}}(Z_{i_2}) - \mu_{\mathbf{X}_i}(Z_{i_2})}{Z_{i_2} - Z_{i_1}} - \frac{\mu_{\mathbf{X}_{i_1}}(Z_{i_1}) - \mu_{\mathbf{X}_i}(Z_{i_1})}{Z_{i_2} - Z_{i_1}} \right| \\
&\leq \frac{1}{2N\kappa\eta} \sum_{i=1}^{N} \{ |\mu_{\mathbf{X}_{i_2}}(Z_{i_2}) - \mu_{\mathbf{X}_i}(Z_{i_2})| + |\mu_{\mathbf{X}_{i_1}}(Z_{i_1}) - \mu_{\mathbf{X}_i}(Z_{i_1})| \} \\
&\leq \frac{C}{2N\kappa\eta} \sum_{i=1}^{N} \{ \|\mathbf{X}_{i_2} - \mathbf{X}_i\| + \|\mathbf{X}_{i_1} - \mathbf{X}_i\| \}
\end{aligned}$$

holds for some constant $C$. Let $N_1$ indicates the number of individuals $j$ having exposure $Z_j$ satisfying $z_0 - \eta \leq Z_j \leq z_0 - \kappa\eta$ and $N_2$ indicates the number of individuals $j$ having exposure $Z_j$ satisfying $z_0 + \kappa\eta \leq Z_j \leq z_0 + \eta$. Then, $N_j = O(N\eta)$ holds for $j = 1, 2$.

Also, Lemma S.1 implies that all the moments of $N_j^{1/d} \|\mathbf{X}_{i_j} - \mathbf{X}_i\|$ is uniformly bounded in $N$ for $j = 1, 2$. Therefore,

$$\begin{aligned}
\left| (N\eta^{d+1})^{1/d} \mathcal{B}_{z_0} \right| &\leq \frac{C}{2N\kappa} \sum_{i=1}^{N} \{ (N\eta)^{1/d} \|\mathbf{X}_{i_2} - \mathbf{X}_i\| + (N\eta)^{1/d} \|\mathbf{X}_{i_1} - \mathbf{X}_i\| \} \\
&= \frac{C}{2N\kappa} \sum_{i=1}^{N} \{ N_2^{1/d} \|\mathbf{X}_{i_2} - \mathbf{X}_i\| + N_1^{1/d} \|\mathbf{X}_{i_1} - \mathbf{X}_i\| \} + O_p(1) \\
&= O_p(1)
\end{aligned}$$

holds. $\qquad\square$

Lemma S.2 highlights the significance of the dimension of covariates in determining the stochastic order of conditional bias. It suggests that as the dimension of covariates increases, the caliper $\eta$ should decrease at a lower rate, which aligns with our chosen setting of $\eta$. In the case where the covariate is scalar, the theorem demonstrates that the bias term is negligible. However, with higher dimensional covariates ($d > 1$), the bias may not be negligible and may impact the asymptotic behaviors of the proposed estimator.

**Lemma S.3** (Number of Replacements). Assume Assumptions 1–4 and S.1 hold. Then, $K_{z_0}(i) = O_p(1/\eta)$ and $\mathbb{E}[\{\eta K_{z_0}(i)\}^q]$ is bounded uniformly in $N$ for any $q > 0$.

*Proof.* Proof follows from the extension of Abadie and Imbens (2006). Also, see Wu et al. (2022). □

Lemma S.3 shows that $\eta K_{z_0}(i)$ is $O_p(1)$ with finite moments, which will be used to show asymptotic properties of matching estimator.

## S.1.3   Proof of Theorem 4

*Proof.* First, $\bar{\delta}(z_0) - \delta(z_0) \overset{p}{\to} 0$ holds from the law of large number. By Lemma S.2, $\mathcal{B}_{z_0} = o_p(1)$ holds. For the residual term,

$$|\mathcal{E}_{z_0}| \leq \frac{1}{2N\kappa\eta} \sum_{i=1}^{N} \{|\varepsilon_{i_2}| + |\varepsilon_{i_1}|\}$$

$$= \frac{1}{2N\kappa\eta} \sum_{i=1}^{N} \{K_{z_0}(i)\,|\varepsilon_i|\}$$

$$\mathbb{E}[N\eta^3 \mathcal{E}_{z_0}^2] \leq \frac{\bar{\sigma}^2(z, \mathbf{x})}{4\kappa^2} \frac{1}{N\eta} \sum_{i=1}^{N} \mathbb{E}[\{\eta K_{z_0}(i)\}^2].$$

By Lemma S.3, $N\eta^3 \mathcal{E}_{z_0}^2 = O_p(1)$ because at most $N_1 + N_2$ terms can be positive in summation and $N_1 + N_2 = O(N\eta)$ holds, which implies $\mathcal{E}_{z_0} = o_p(1)$. Next, we will show

that the approximation term also converges to 0 in probability.

$$\frac{\mu_{\mathbf{X}_i}(Z_{i_2}) - \mu_{\mathbf{X}_i}(Z_{i_1})}{Z_{i_2} - Z_{i_1}} = \mu'_{\mathbf{X}_i}(Z_i^*)$$

holds for some $Z_i^*$ between $Z_{i_1}$ and $Z_{i_2}$ by mean value theorem. Also,

$$
\begin{aligned}
\left|\eta^{-1}\mathcal{A}_{z_0}\right| &\leq \frac{1}{N\eta} \sum_{i=1}^{N} \left|\frac{\mu_{\mathbf{X}_i}(Z_{i_2}) - \mu_{\mathbf{X}_i}(Z_{i_1})}{Z_{i_2} - Z_{i_1}} - \mu'_{\mathbf{X}_i}(z_0)\right| \\
&= \frac{1}{N\eta} \sum_{i=1}^{N} \left|\mu'_{\mathbf{X}_i}(Z_i^*) - \mu'_{\mathbf{X}_i}(z_0)\right| \\
&= \frac{1}{N\eta} \sum_{i=1}^{N} \left|\mu''_{\mathbf{X}_i}(Z_i^{**})\right| \left|Z_i^* - z_0\right|
\end{aligned}
$$

holds for some $Z_i^{**}$ between $Z_i^*$ and $z_0$. Since $\mu''_{\mathbf{X}_i}$ is uniformly bounded and $|Z_i^* - z_0| \leq \eta$, $\mathcal{A}_{z_0} = o_p(1)$. Since $\hat{\delta}(z_0) - \delta(z_0) = (\bar{\delta}(z_0) - \delta(z_0)) + \mathcal{E}_{z_0} + \mathcal{A}_{t_0} + \mathcal{B}_{t_0}$ and four terms together converge to 0 in probability, Theorem 4 holds. $\qquad\square$

## S.1.4  Proof of Theorem 5

*Proof.* We begin by proving that $\Sigma_{z_0}^{-1/2} = O_p(1)$. By the Cauchy-Schwartz inequality,

$$\sum_{j=1}^{N} L_{z_0}(j)^2 \sigma_j^2 I_j \geq \frac{1}{N_1 + N_2} \left(\sum_{j=1}^{N} |L_{z_0}(j)\sigma_j I_j|\right)^2$$

holds.

Therefore, $\Sigma_{z_0}^{-1/2}$ can be represented as

$$
\begin{aligned}
\Sigma_{z_0}^{-1/2} &= \left\{\frac{N}{\eta^3 (\sum_{j=1}^{N} L_{z_0}(j)^2 \sigma_j^2 I_j)}\right\}^{1/2} \\
&\leq \left\{\frac{N}{\frac{\eta}{N_1 + N_2}(\sum_{j=1}^{N} |\eta L_{z_0}(j)\sigma_j I_j|)^2}\right\}^{1/2} \\
&\leq \frac{1}{\underline{\sigma}(z,\mathbf{x})\frac{1}{N}\sum_{j=1}^{N} |\eta L_{z_0}(j) I_j|} \times \left(\frac{N_1 + N_2}{N\eta}\right)^{1/2}.
\end{aligned}
$$

5

Also, since $N_1 + N_2 = O(N\eta)$ and $\frac{K_{z_0}(j)}{2\eta} \leq |L_{z_0}(j)| \leq \frac{K_{z_0}(j)}{2\kappa\eta}$ hold,

$$\frac{1}{N}\sum_{j=1}^{N}|\eta L_{z_0}(j)I_j| \geq \frac{1}{2N}\sum_{j=1}^{N}K_{z_0}(j)I_j = 1,$$

and thus, $\Sigma_{z_0}^{-1/2} = O_p(1)$.

Recall that we can decompose

$$\hat{\delta}(z_0) - \mathcal{B}_{z_0} - \mathcal{A}_{z_0} - \delta(z_0) = (\bar{\delta}(z_0) - \delta(z_0)) + \mathcal{E}_{z_0}.$$

By the standard central limit theorem,

$$N^{1/2}\{\bar{\delta}(z_0) - \delta(z_0)\} \xrightarrow{d} N(0, \mathbb{E}[\{\mu'_{\mathbf{X}_i}(z_0) - \mu'(z_0)\}^2])$$

where $\mu(z) = \mathbb{E}[Y_i(z)]$. Boundedness of $\mu'_{\mathbf{X}_i}(z_0)$ guarantees the boundedness of asymptotic variance. Considering the fact that $\Sigma_{z_0}^{-1/2} = O_p(1)$, $\Sigma_{z_0}^{-1/2}(N\eta^3)^{1/2}(\bar{\delta}(z_0) - \delta(z_0)) = o_p(1)$ also holds.

For the last term $\Sigma_{z_0}^{-1/2}(N\eta^3)^{1/2}\mathcal{E}_{z_0}$, we will see the distribution of

$$\frac{(N\eta^3)^{1/2}\mathcal{E}_{z_0}}{\Sigma_{z_0}^{1/2}} = \frac{\sum_{j=1}^{N}L_{z_0}(j)\varepsilon_j I_j}{(\sum_{j=1}^{N}L_{z_0}(j)^2\sigma_j^2 I_j)^{1/2}}.$$

Note that $\varepsilon_j$ are independent with zero means and variances $\sigma_j^2$ conditioning on $(Z_j, \mathbf{X}_j)$. We will use the Lindeberg-Feller central limit theorem to show its distribution. We verify the Lyapunov condition to use the Lindeberg-Feller central limit theorem. Conditioning on $(Z_j, \mathbf{X}_j)$, the Lyapunov condition is

$$\frac{\sum_{j=1}^{N}\mathbb{E}[|L_{z_0}(j)\varepsilon_j I_j|^{2+\kappa}|Z_j, \mathbf{X}_j]}{(\sum_{j=1}^{N}L_{z_0}(j)^2\sigma_j^2 I_j)^{\frac{2+\kappa}{2}}} = \frac{\sum_{j=1}^{N}\mathbb{E}[|\eta^2 L_{z_0}(j)\varepsilon_j I_j|^{2+\kappa}|Z_j, \mathbf{X}_j]}{(\sum_{j=1}^{N}\eta^4 L_{z_0}(j)^2\sigma_j^2 I_j)^{\frac{2+\kappa}{2}}}$$
$$\leq \frac{\sum_{j=1}^{N}\mathbb{E}[|\eta^2 L_{z_0}(j)\varepsilon_j I_j|^{2+\kappa}|Z_j, \mathbf{X}_j]}{\{\frac{1}{N_1+N_2}(\sum_{j=1}^{N}|\eta^2 L_{z_0}(j)\sigma_j I_j|)^2\}^{\frac{2+\kappa}{2}}}.$$

6

The latter inequality holds by the Cauchy-Schwartz inequality.

$$\sum_{j=1}^{N} \eta^4 L_{z_0}(j)^2 \sigma_j^2 I_j \geq \frac{1}{N_1 + N_2} (\sum_{j=1}^{N} |\eta^2 L_{z_0}(j)\sigma_j I_j|)^2.$$

Also, let $C_\kappa = \sup_{z, \mathbf{x}} \mathbb{E}[|Y_j - \mu(Z_j, \mathbf{X}_j)|^{2+\kappa} | Z_j = z, \mathbf{X}_j = \mathbf{x}]$. By Assumption S. 1, $C_\kappa$ is uniformly bounded. Therefore, the following holds.

$$\frac{\sum_{j=1}^{N} \mathbb{E}[|\eta^2 L_{z_0}(j)\varepsilon_j I_j|^{2+\kappa} | Z_j, \mathbf{X}_j]}{\{\frac{1}{N_1+N_2}(\sum_{j=1}^{N} |\eta^2 L_{z_0}(j)\sigma_j I_j|)^2\}^{\frac{2+\kappa}{2}}} \leq \frac{(N_1 + N_2)C_\kappa \mathbb{E}[|\eta^2 L_{z_0}(j)|^{2+\kappa}]}{(N_1 + N_2)^{-\frac{2+\kappa}{2}} \eta^{2+\kappa} \underline{\sigma}^{2+\kappa}(z, \mathbf{x})(\sum_{j=1}^{N} |\eta L_{z_0}(j) I_j|)^{2+\kappa}}$$

$$= \left(\frac{N_1 + N_2}{N\eta}\right)^{\frac{4+\kappa}{2}} \frac{C_\kappa}{\underline{\sigma}^{2+\kappa}(z, \mathbf{x})(N\eta)^{\kappa/2}} \frac{\mathbb{E}[|\eta^2 L_{z_0}(j)|^{2+\kappa}]}{(\frac{1}{N}\sum_{j=1}^{N} |\eta L_{z_0}(j) I_j|)^{2+\kappa}}.$$

Since $\frac{N_1 + N_2}{N\eta} = O(1)$ and $(N\eta)^{\kappa/2} \to \infty$, it is enough to show that the latter term is bounded. Recall that $\frac{K_{z_0}(j)}{2\eta} \leq |L_{z_0}(j)| \leq \frac{K_{z_0}(j)}{2\kappa\eta}$. The numerator of the last term can be bounded as

$$\mathbb{E}[|\eta^2 L_{z_0}(j)|^{2+\kappa}] \leq \frac{1}{(2\kappa)^{2+\kappa}} \mathbb{E}[(\eta K_{z_0}(j))^{2+\kappa}]$$

by Lemma S.3. Also, the denominator can be expressed as

$$(\frac{1}{N}\sum_{j=1}^{N} |\eta L_{z_0}(j) I_j|)^{2+\kappa} \geq \frac{1}{2^{2+\kappa}} \left(\frac{1}{N}\sum_{j=1}^{N} K_{z_0}(j) I_j\right)^{2+\kappa} = 1.$$

Thus, the Lyapunov condition is satisfied for almost all $(Z_j, \mathbf{X}_j)$, which implies the Lindberg-Feller central limit theorem.

$$\Sigma_{z_0}^{-1/2} (N\eta^3)^{1/2} \mathcal{E}_{z_0} \xrightarrow{d} \mathcal{N}(0, 1).$$

$\square$

### S.1.5  Proof of Corollary 1

*Proof.* Assume that $\mathbf{X}_i$ is a scalar. Then, $\mathcal{A}_{z_0} = O_p(\eta)$ from the proof of Theorem 4, and Lemma S.2 implies $\mathcal{B}_{z_0} = O_p\{(N\eta^2)^{-1}\}$. Therefore, $\Sigma_{z_0}^{-1/2}(N\eta^3)^{1/2}(\mathcal{A}_{z_0} + \mathcal{B}_{z_0}) = o_p(1)$ considering the fact that $\eta = o(N^{-1/4})$ and $\Sigma_{z_0}^{-1/2} = O_p(1)$. $\qquad\square$

## S.2  Extension to 1:$M$ matching

The matching algorithm we proposed can be extended to 1:$M$ matching. The only difference is that, for each $i$, we choose $M$ units $\{i_{1,1}, \ldots, i_{1,M}\}$ in $[z_0 - \eta, z_0 - \kappa\eta]$ and $\{i_{2,1}, \ldots, i_{2,M}\}$ in $[z_0 + \kappa\eta, z_0 + \eta]$, respectively. Then, we may choose either the mean or median from $M \times M$ possible combinations of $(Y_{i_{2,m}} - Y_{i_{1,m}})/(Z_{i_{2,m}} - Z_{i_{1,m}})$ as an estimator for individual derivative effect of $i$. To summarize, we modify Algorithm 1 in the paper as Algorithm S.1.

**Algorithm S.1** $M \times M$ Matching Algorithm for Average Causal Derivative Effect

**Input:** Data $\mathcal{D} = \{(\mathbf{x}_i, y_i, z_i)\}_{i=1}^N$, a metric $d(\cdot, \cdot)$, an exposure level of interest $z_0 \in \mathbb{Z}$, blocks of $Z$: $Z_k$, $k = 1, \ldots, K$, and the number of matched units $M$.

**Output:** Matched set $\{(i_{1,1}, \ldots, i_{1,M}), (i_{2,1}, \ldots, i_{2,M})\}_{i=1}^N$ and an estimated causal derivative effect at $z_0$, $\hat{\delta}(z_0)$.

(a) Specify a candidate set of $(\eta, \kappa)$ with small $\eta(> 0)$ values and $\kappa$ ranging from 0 to 1, or fix $\kappa$ to 0.1 by the rule of thumb.

(b) For each pair of $(\eta, \kappa)$ from the candidate set,

**for** $i = 1, 2, \ldots, N$ **do**

   Select $i_{1,m}$ and $i_{2,m}$, $m = 1, \ldots, M$ from the set $\{1, 2, \ldots, N\}$ satisfying

$$i_{1,m} = \sum_{l:z_0-\eta \leq Z_l \leq z_0-\kappa\eta} \mathbb{I}\{d(\mathbf{X}_l, \mathbf{X}_i) \leq d(\mathbf{X}_{i_{1,m}}, \mathbf{X}_i)\} = m,$$

$$i_{2,m} = \sum_{l:z_0+\kappa\eta \leq Z_l \leq z_0+\eta} \mathbb{I}\{d(\mathbf{X}_l, \mathbf{X}_i) \leq d(\mathbf{X}_{i_{2,m}}, \mathbf{X}_i)\} = m.$$

   That is, within each interval, we select the $m$th closest unit to unit $i$ based on covariate distances.

**end for**

(c) Calculate the average BASMD by (3) where $\mathbf{X}^* = \{\mathbf{X}_1, \ldots, \mathbf{X}_N\}$, $\mathbf{X}_1^* = \{\mathbf{X}_{1,1}, \ldots, \mathbf{X}_{1,M} \ldots, \mathbf{X}_{N,1}, \ldots, \mathbf{X}_{N,M}\}$, $\mathbf{X}_2^* = \{\mathbf{X}_{1,1}, \ldots, \mathbf{X}_{1,M} \ldots, \mathbf{X}_{N,1}, \ldots, \mathbf{X}_{N,M}\}$ for each matched dataset from step (b).

(d) Choose the optimal $(\eta, \kappa)$ that minimizes the average BASMD and determine the corresponding matched set $\{(i_{1,1}, \ldots, i_{1,M}), (i_{2,1}, \ldots, i_{2,M})\}_{i=1}^N$, ensuring the best covariate balance.

(e) Using the matched set $\{(i_{1,1}, \ldots, i_{1,M}), (i_{2,1}, \ldots, i_{2,M})\}_{i=1}^N$ in (d), estimate the ACDE at $z_0$ as $\hat{\delta}(z_0) = \frac{1}{N}\sum_{i=1}^N \hat{\delta}_i(z_0)$ where $\hat{\delta}_i(z_0)$ can be either the mean or median from $M \times M$ possible combinations of $(Y_{i_{2,m}} - Y_{i_{1,m}})/(Z_{i_{2,m}} - Z_{i_{1,m}})$.

## S.3 Covariate Adjustment

In this section, we briefly introduce a covariate adjustment method that can be employed to mitigate bias from unbalanced covariates. Rosenbaum (2002) applied covariate adjustment in the estimation, the testing hypothesis procedure, and sensitivity analysis. If the covariates for matched units exhibit significant imbalances, the inference stage is significantly affected by these covariate differences. Therefore, it becomes necessary to adjust these covariates to ensure the validity of inferences. In our settings, we employ a similar method to Rosenbaum (2002). The main idea is to use residuals instead of the outcome itself in the analysis. Let $\varepsilon(\cdot)$ represent the function that generates residuals from $Y$ and $\mathbf{X}$. Specifically, we choose to fit $Y$ using $\mathbf{X}$ through least squares linear regression, $Y = \mathbf{X}\beta + \epsilon$, where $\varepsilon(Y_i) = Y_i - \mathbf{X}_i\hat{\beta}$. Utilizing $\varepsilon(Y_i)$ instead of $Y_i$ in the analysis allows us to mitigate bias induced by unbalanced covariates, as we have adjusted for the covariate imbalances.

Note that since we do not use the outcome variable in the matching stage, the matching structure and optimal parameters remain the same in the original method and covariate adjustment method. However, we estimate the ACDE at $z_0$ as:

$$\hat{\delta}(z_0) = \frac{1}{N} \sum_{i=1}^{N} \frac{\varepsilon(Y_{i_2}) - \varepsilon(Y_{i_1})}{Z_{i_2} - Z_{i_1}}$$
$$= \frac{1}{N} \sum_{i=1}^{N} \frac{(Y_{i_2} - Y_{i_1}) - (\mathbf{X}_{i_2} - \mathbf{X}_{i_1})\hat{\beta}}{Z_{i_2} - Z_{i_1}}.$$

If there is an exact match on the covariate, this method gives the same result. Similarly, we use the residuals $\varepsilon(Y_i)$ instead of the outcomes $Y_i$ in the testing hypothesis procedure and sensitivity analysis. We apply the covariate adjustment method to the COPD data in Section 6. Tables S.1 and S.2 show the results when we adjust the covariate.

Table S.1: Estimation of the ACDE of emphysema is conducted at each level of fSAD after covariate adjustment. The optimal values for $\eta$ and $\kappa$, determined based on the average BASMD, are reported. Additionally, p-values are calculated using the permutation test.

| fSAD | $\eta$ | $\kappa$ | Average BASMD | ACDE | p-value |
|------|--------|----------|---------------|------|---------|
| 0.04 | 0.07 | 0.2 | 0.413 | -0.055 | 0.734 |
| **0.06** | 0.08 | 0.1 | 0.334 | **0.235** | **0.002** |
| **0.08** | 0.08 | 0.15 | 0.170 | **0.288** | **0.001** |
| 0.10 | 0.08 | 0.15 | 0.172 | 0.066 | 0.238 |
| 0.12 | 0.075 | 0.15 | 0.122 | 0.015 | 0.843 |
| 0.14 | 0.075 | 0.1 | 0.138 | 0.035 | 0.790 |
| 0.16 | 0.055 | 0.1 | 0.140 | -0.222 | 0.300 |
| **0.18** | 0.075 | 0.1 | 0.083 | **-0.284** | **0.020** |
| 0.20 | 0.065 | 0.15 | 0.107 | -0.316 | 0.064 |

Table S.2: Sensitivity analysis for unmeasured confounders after covariate adjustments at fSAD levels of 0.08 and 0.18.

| $\Gamma$ | fSAD = 0.08 | | fSAD = 0.18 | |
|----------|-------------|-------------|-------------|-------------|
| | Lower p-value | Upper p-value | Lower p-value | Upper p-value |
| 1.000 | **0.001** | **0.001** | **0.016** | **0.016** |
| 1.050 | **< 0.001** | **0.003** | **0.009** | **0.028** |
| 1.100 | **< 0.001** | **0.008** | **0.005** | **0.045** |
| 1.112 | **< 0.001** | **0.009** | **0.004** | **0.050** |
| 1.150 | **< 0.001** | **0.015** | **0.003** | 0.069 |
| 1.200 | **< 0.001** | **0.028** | **0.001** | 0.100 |
| 1.250 | **< 0.001** | **0.048** | **< 0.001** | 0.140 |
| 1.254 | **< 0.001** | **0.050** | **< 0.001** | 0.144 |



\end{document}